\documentclass[12pt]{iopart}

\usepackage{amssymb}
\usepackage{amsbsy}

\newcommand{\be}{\begin{equation}} \newcommand{\ee}{\end{equation}}
\newcommand{\bea}{\begin{eqnarray}} \newcommand{\eea}{\end{eqnarray}}
\newcommand{\el}{\nonumber \\}
\newcommand{\re}[1]{(\ref{#1})}
\newcommand{\pat}{\partial}

\newcommand{\Rdot}{\dot{R}}

\newcommand{\brt}[1]{[#1]}
\renewcommand{\t}{^2}

\newcommand{\PRD}[1]{{\it Phys. Rev.} {\bf D#1}}
\renewcommand{\PRL}[1]{Phys. Rev. Lett. {\bf #1}}

\newcommand{\NPB}[1]{{\it Nucl. Phys.} {\bf B#1}}

\newcommand{\MNRAS}[1]{{\it Mon. Not. Roy. Astron. Soc.} {\bf #1}}
\newcommand{\APJ}[1]{{\it Astrophys. J.} {\bf #1}}

\renewcommand{\CQG}[1]{Class. Quant. Grav. {\bf #1}}
\newcommand{\GRG}[1]{{\it Gen. Rel. Grav.} {\bf #1}}
\renewcommand{\AA}[1]{{\it Astron. \& Astrophys.} {\bf #1}}

\begin{document}


\title{Backreaction in the Lema\^{\i}tre-Tolman-Bondi model}

\author{Syksy R\"{a}s\"{a}nen}

\address{Theoretical Physics, University of Oxford,
1 Keble Road, Oxford, OX1 3NP, UK}

\ead{syksy.rasanen@iki.fi}

\begin{abstract}

\noindent We study backreaction analytically using the
parabolic Lema\^{\i}tre-Tolman-Bondi universe as a toy model.
We calculate the average expansion rate and energy density
on two different hypersurfaces and compare the results.
We also consider the Hubble law and find that backreaction
slows down the expansion if measured with proper time, but
speeds it up if measured with energy density.

\end{abstract}



\setcounter{secnumdepth}{3}

\section{Introduction}

The homogeneous and isotropic Friedmann-Robertson-Walker models
are usually thought to give a good description of the average
behaviour of the universe, since the universe appears to be
homogeneous and isotropic on sufficiently large
scales (though see \cite{Pietronero:2004}).
The reasoning behind the FRW equations is that one takes
the average of the real (inhomogeneous and anisotropic)
metric and energy-momentum tensor, and plugs these into the
Einstein equation. However, the physically correct thing would
be to first plug the real inhomogeneous and anisotropic
quantities into the Einstein equation, and only then take the
average. Because the Einstein equation is non-linear, these
two procedures are not equivalent. In other words, the averages
of the real quantities do not satisfy the Einstein equation.
The feature that the average behaviour of an
inhomogeneous spacetime is not the same as the behaviour
of the corresponding smooth spacetime (that is, one with the
same average initial conditions) is called backreaction.
In the context of cosmology, the issue was highlighted in
\cite{Ellis} (it had been discussed earlier in \cite{Shirokov:1963})
as the \emph{fitting problem:} how do we find the homogeneous and
isotropic model which best fits the real inhomogeneous and anisotropic
universe?

The relation between the average sources (energy density, pressure, etc.) 
and average geometric quantities (expansion rate, shear, etc.) in a
general spacetime filled with a perfect irrotational fluid
is known \cite{Buchert:1999a, Buchert:2001}. However, the system of
equations is not closed, which means that different inhomogeneous
spacetimes with the same initial averages
evolve differently even as far as average
quantities are concerned. Since there is no procedure for
finding the average behaviour of a given spacetime (short
of solving it exactly), it is difficult to quantitatively
evaluate the importance of backreaction in cosmology. One
notable calculation is \cite{Buchert:1999b}, where the
behaviour of an universe-sized box filled with
inhomogeneous expanding dust is followed numerically
using the Newtonian limit of the exact backreaction formalism
of \cite{Buchert:1999a, Buchert:1995}. The results show
unambiguously that backreaction is a real phenomenon which can have
a large impact: for example, it can make regions with
no initial overdensity turn around and collapse. However,
the Newtonian limit is not expected to fully capture the
relativistic backreaction. (In particular, the global backreaction
vanishes identically for periodic boundary conditions, unlike
in the relativistic case.)

Fully relativistic quantitative studies of backreaction have 
usually been done in the context of perturbative solutions around
a FRW background
\cite{Mukhanov, Wetterich:2001, Geshnizjani:2002, Finelli:2003, Geshnizjani:2003, Rasanen:2003, Brandenberger:2004a, Geshnizjani:2004, Rasanen:2004, Brandenberger:2004b}
(see \cite{Rasanen:2003} for a more extensive list of references).
Of particular interest has been the possibility that
backreaction of long wavelength perturbations
could lead to a dynamical relaxation of vacuum energy,
at the same stroke providing an elegant inflationary
mechanism and explaining why the vacuum energy
is so much smaller than theoretically expected
\cite{Brandenberger:2004a, Brandenberger:2004b, Tsamis, Brandenberger:2000, Brandenberger:2002}. It has also been suggested that this could solve the
coincidence problem of the effective vacuum energy density
being of the order of the matter energy density today
\cite{Brandenberger:2000, Brandenberger:2002}. Another possible
explanation for the coincidence problem is backreaction from
structure formation, that is, from small wavelength modes
\cite{Wetterich:2001, Rasanen:2003, Rasanen:2004, Schwarz:2002}.
Since the acceleration of the universe seems to have started around the
era when structure formation is important, it seems a natural
possibility that this deviation from the simple prediction of
deceleration in the homogeneous and isotropic models
with normal matter could be related
to the growth of inhomogeneities in the universe.

Perturbative treatment of backreaction has two drawbacks.
First, naive perturbative results break down when
perturbations have an effect on the background, in other words
when backreaction is important. In addition to the usual issues
of cosmological perturbation theory such as gauge-invariance
\cite{Mukhanov, Finelli:2003, Unruh:1998}, one has to worry about
new problems such as convergence and consistency of the perturbative
expansion. These make consistent backreaction
calculations in the perturbative framework an involved task.
Second, present-day universe is not perturbatively close
to homogeneity, but contains non-linear structures. If
the coincidence problem is to be solved by backreaction from
structure formation, one would have to go beyond
perturbation theory (though if the effect involves mainly
perturbations breaking away from the linear regime,
a quasi-perturbative treatment might still be possible).

We will study backreaction analytically and free of
perturbative ambiguities with an exact toy model, 
the Lema\^{\i}tre-Tolman-Bondi model
\cite{Lemaitre:1933, Tolman:1934, Bondi:1947}
(see also \cite{Krasinski:1997, Sussman:2001}).
The LTB model is the spherically symmetric dust
solution of the Einstein equation (or rather the
family of such solutions). Like the perturbed FRW
spacetime, the LTB spacetime is a generalisation of
the FRW universe. The LTB model can be viewed as an
Einstein-de Sitter universe with a single spherically
symmetric perturbation which can be arbitrarily large, as
opposed to the linearly perturbed FRW universe
which has an ensemble of small perturbations. The LTB
model can describe the collapse of an overdensity
or the formation of a void in an expanding universe
\cite{Krasinski}, and it has also been used to
describe the local inhomogeneous universe \cite{clusters}.
As a model for the entire universe, it is a toy
model which is useful because the backreaction problem
can be studied quantitatively and without any approximations,
as has been previously done in \cite{Hellaby:1988} (averaging
in the LTB model has also been discussed in \cite{Mustapha:1997};
see also \cite{Buchert:2003}).

In a perturbative framework one aspect of backreaction
is that inhomogeneities change the behaviour of a smooth
background. This feature is obviously not present in
an exact inhomogeneous solution, but the difference between
the average behaviour and the behaviour of
the corresponding smooth spacetime
can be studied unambiguously. With an exact solution one
can also study the choice of the hypersurface of averaging
in isolation of the different issue of the choice of gauge
\cite{Mukhanov, Geshnizjani:2002, Finelli:2003, Geshnizjani:2003, Rasanen:2003, Unruh:1998}.

The present study is a simple exact non-perturbative
counterpart of \cite{Rasanen:2003}, where we calculated
perturbatively the average expansion rate of a linearly
perturbed Einstein-de Sitter universe.
In section 2 we calculate the average expansion rate
in the simplest LTB solution for two different choices of
hypersurface, and compare the results with each other and the FRW universe.
In section 3 we look at the impact of backreaction on the
Hubble law and discuss our results.

\section{The backreaction calculation}

\subsection{The Lema\^{\i}tre-Tolman-Bondi model}

\paragraph{The metric.}

We will proceed as in \cite{Rasanen:2003}: we will write down the
metric, find the proper time and the
expansion rate and take the average over the hypersurface of
proper time. We will also take the average over the hypersurface
of constant coordinate time, and compare the results.

The LTB model is the most general spherically symmetric dust
solution. The Einstein equation is (we take the cosmological
constant to be zero)
\bea \label{einstein1}
  G_{\mu\nu} = \frac{1}{M_{Pl}^2} T_{\mu\nu} = \frac{1}{M_{Pl}^2} \rho u_{\mu} u_{\nu} \ ,
\eea

\noindent where $M_{Pl}=1/\sqrt{8 \pi G_N}$ is the (reduced)
Planck mass, $\rho$ is the energy density and $u^{\mu}$ is the
velocity of the matter fluid, with $u_{\mu} u^{\mu}=-1$.
The spherical symmetry allows the Einstein equation \re{einstein1}
to be solved exactly. The metric turns out to be
\bea \label{metric1}
  \rmd s^2 = - \rmd t^2 + \frac{R'(t,r)^2}{1+E(r)} \rmd r^2 + R(t,r)^2  (\rmd\theta^2 + \sin^2\theta\rmd\phi^2) \ ,
\eea

\noindent where the functions $R(t,r)$ and $E(r)$ are related to
each other and to the energy density $\rho(t,r)$ as follows
\bea \label{einstein2}
  \Rdot(t,r)^2 &=& \frac{1}{M_{Pl}^2} \frac{m(r)}{R(t,r)} + E(r) \el
  \rho(t,r) &=& \frac{m'(r)}{R(t,r)^2 R'(t,r)} \ ,
\eea

\noindent where dots and primes denote derivatives with respect to
$t$ and $r$, respectively, and $m(r)$ is a function which describes
how much energy there is within the radius $r$. (The
freedom to choose the $r$-coordinate means that $m(r)$ can be
redefined at will.) The velocity of the matter fluid is simply
\bea \label{u}
  u^{\mu} = (1,0,0,0) \ . 
\eea

There are three different classes of solutions to
\re{einstein2}, for $E>0, E<0$ and $E=0$. We will
consider the solution with $E=0$, known as
the parabolic solution, which is the simplest and
the most analogous to the Einstein-de Sitter universe.
We assume that $m'>0$ and choose the radial coordinate such that
$m(r)=4 M_{Pl}^2 r^3/(9 t_1^2)$, where $t_1$ is a positive
constant with the dimension of time. We then have
\bea \label{R}
  R(t,r) = r \left( \frac{t-t_0(r)}{t_1} \right)^{\frac{2}{3}} \ ,
\eea

\noindent where $t_0(r)$ is an arbitrary function. It is
transparent that the solution reduces to the FRW case if
$t_0(r)=$ constant. There is a singularity at $t=t_0(r)$,
and we have chosen to look at the case $t>t_0(r)$,
which describes matter expanding away from the big bang
rather than collapsing towards a future singularity.
The function $t_0(r)$ is the big bang time
for a comoving observer at constant $r$: different observers
find their part of the universe to have emerged from the singularity at
different times in the past. In order to avoid a singularity
caused by shells of matter crossing, we must have $t_0'(r)\leq0$
\cite{Hellaby:1985}.

\paragraph{The proper time.}

As emphasised in \cite{Geshnizjani:2002, Rasanen:2003, Rasanen:2004}, it is
important to cast things in terms of the proper time of the observer.
Given \re{u}, the derivative in the direction orthogonal to
the hypersurface defined by the velocity of comoving observers
is $\pat_{\tau}=u^{\mu}\nabla_{\mu}=\nabla_t$.
From the condition $\pat_{\tau}\tau=1$ we get the proper
time $\tau = t + f(r)$, where $f(r)$ is an arbitrary function.

In the perturbed FRW case \cite{Rasanen:2003} there was a
difference between the background coordinate time $t$ and
the proper time $\tau$ because the velocity $u^{\mu}$ of the
perturbed comoving observers was different from the background
velocity, leading to a perturbed $\tau$. Here the situation is different.
The velocity of comoving observers is the same as in the smooth
case, but different observers start their clocks at different times.
When discussing the expansion rate measured by cosmological
observers, we want to compare observers whose clocks show the same
time. We therefore choose $f(r)=-t_0(r)$, so the proper time is
\bea \label{tau}
  \tau(t,r) = t - t_0(r) \ ,
\eea

\noindent which is the ``local age of the universe'', the time
since the big bang measured by a comoving observer at constant $r$.
Note that in order for $\tau$ to be a time coordinate, the normal
$\nabla_{\mu}\tau$ has to be timelike, which translates into the
constraint $R'(t,r)>-t_0'(r)$.

\paragraph{The expansion rate.}

We are interested in the expansion rate measured by a
comoving observer, given by
\bea \label{thetadef}
  \theta(t,r) = \nabla_{\mu} u^{\mu} \ ,
\eea

\noindent which in the FRW case reduces to $\theta=3 H$, where $H$ 
is the Hubble parameter. Plugging in \re{metric1}, \re{u} and \re{R} we have
\bea \label{theta}
  \theta(t,r) = \frac{2}{\tau} \left( 1 - \frac{r\tau'}{3\tau + 2 r\tau'} \right) \ .
\eea

The expansion rate of an Einstein-de Sitter universe is
$2/\tau$, so the second term is the contribution of the
inhomogeneity. Since $\tau'=-t_0'>0$, this term is negative,
and the expansion in terms of the proper time
is always slower than in the FRW case, and slows
down with increasing $r$. In the converse situation with $\tau'<0$,
the expansion rate would be larger than in the FRW case,
but the inner shells of matter would overtake the outer
shells, resulting in a singularity. (In a more realistic model,
shell-crossing would mean that the decription of matter as a
pressureless ideal fluid breaks down, not necessarily that there
is a singularity.)

\paragraph{The energy density.}

From \re{einstein2} and \re{R} we have
\bea \label{rho}
  \rho(t,r) = \frac{4 M_{Pl}^2}{3\tau^2} \frac{3\tau}{3\tau + 2 r\tau'} \ .
\eea

In the solution with $E=0$, the density is uniquely determined
by the local big bang time $t_0$, and one can see from $\rho$
concretely what kind of a physical situation a given $t_0$
corresponds to. In particular, when $t_0$ is constant, \re{rho}
reduces to the FRW case, $\rho=4 M_{Pl}^2/(3\tau^2)$. It is
possible to specify a LTB solution by giving the energy density
on initial and final hypersurfaces \cite{Krasinski}, instead
of giving $E$ and $t_0$ as done here.

\subsection{Taking the average}

\paragraph{The average expansion rate.}

To obtain the average expansion rate, we should integrate the
local expansion rate \re{theta} over the appropriate hypersurface.
To find the integration measure on the hypersurface of constant
$\tau$, let us write the metric \re{metric1} in terms of $\tau$:
\bea \label{metric2}
  \fl \rmd s^2 &=& - \rmd\tau^2 - 2 t_0'\rmd\tau\rmd r + (R'^2 -t_0'^2) \rmd r^2 + R^2 (\rmd\theta^2 + \sin^2\theta\rmd\phi^2) \el
  \fl &=& - \frac{R'^2}{R'^2-t_0'^2}\rmd\tau^2 + (R'^2 -t_0'^2) \left( \rmd r - \frac{t_0'}{R'^2-t_0'^2}\rmd\tau \right)^2 + R^2 (\rmd\theta^2 + \sin^2\theta\rmd\phi^2) \ ,
\eea

\noindent where we again see that $R'>-t_0'$ is required
for $\tau$ to be a time coordinate.  We can now read off the
integration measure on the hypersurface of constant $\tau$ \cite{Misner:1970}:
\bea \label{measure}
  |\det{}^{(\tau)}g_{ij}|^{1/2} = (R'^2 - t_0'^2)^{1/2} R^2\sin\theta \ .
\eea

The average of a scalar observable $F(t,r)$ over the hypersurface of
constant $\tau$ is then
\bea \label{averagetau}
  \langle F \rangle_{\tau_c} &=& \frac{\int\rmd^4 x |\det{}^{(\tau)}g_{ij}|^{1/2} \delta(t-t_0(r)-\tau_c) F(t,r)}{\int\rmd^4 x |\det{}^{(\tau)}g_{ij}|^{1/2} \delta(t-t_0(r)-\tau_c)} \el
  &=& \frac{\int_0^{\infty}\rmd r r\t \left[ \left(\frac{\tau_c}{t_1}\right)^{\frac{4}{3}} \left(\frac{3\tau_c - 2 r t_0'}{3\tau_c}\right)\t - t_0'\,\t \right]^{1/2} F(t_0(r)+\tau_c,r)}{\int_0^{\infty}\rmd r r\t \left[ \left(\frac{\tau_c}{t_1}\right)^{\frac{4}{3}} \left(\frac{3\tau_c - 2 r t_0'}{3\tau_c}\right)\t - t_0'\,\t \right]^{1/2}} \ ,
\eea

\noindent where $\delta(t-t_0(r)-\tau_c)$ is the delta function,
$\tau_c$ is the constant value of proper time which labels the
hypersurface, and on the second line we have plugged in \re{R}
and \re{measure}. To keep the analogy with the FRW case as close
as possible, we will consider only spacetimes where the
coordinate $r$ ranges from 0 to $\infty$.

In comparison, the integration measure on the hypersurface of
constant $t$ is
\bea \label{measuret}
  |\det{}^{(t)}g_{ij}|^{1/2} = R' R^2\sin\theta \ ,
\eea

\noindent and the average over the hypersurface of constant $t$ reads
\bea \label{averaget}
  \langle F \rangle_{t_c} &=& \frac{\int\rmd^4 x |\det{}^{(t)}g_{ij}|^{1/2} \delta(t-t_c) F(t,r)}{\int\rmd^4 x |\det{}^{(t)}g_{ij}|^{1/2} \delta(t-t_c)} \el
  &=& \frac{ \int_0^{\infty}\rmd r r\t (t_c-t_0) [3 (t_c-t_0) - 2 r t_0'] F(t_c,r) }{ \int_0^{\infty}\rmd r r\t (t_c-t_0) [3 (t_c-t_0) - 2 r t_0'] } \ ,
\eea

\noindent where $t_c$ is again the constant labelling the hypersurface.
There are two differences between the averages
\re{averagetau} and \re{averaget}. The integration measure is different,
and the quantity being held constant in the integral is different.
In general, both are expected to be important, but in the present
case the integration measure will in fact turn out not to matter.
We will now evaluate the average of the expansion rate
$\theta$ over both hypersurfaces for some simple choices of $t_0$.

\paragraph{Comparing the averages.}

Our LTB solution is specified once we give the function
$t_0$, which has to satisfy $R'>-t_0'\geq0$. We consider
three examples of $t_0$. They are given in Table \ref{table:averages},
along with the average expansion rates and average densities
($r_0$ is a positive constant).
The exponential factors in the first two cases are needed for satisfying the
condition $R'>-t_0'$; without them, the averages would be the same but
the hypersurfaces would not be spacelike for all $\tau_c$.
We also give the effective scale factor $a_{\tau}$ defined
by $3\pat_{\tau}a_{\tau}/a_{\tau}=\langle\theta\rangle_{\tau}$, in analogy with
\cite{Buchert:1999a, Buchert:2001, Buchert:1999b, Buchert:1995}\footnote{Note that this
definition is different from the one used in \cite{Rasanen:2003, Rasanen:2004}.}
($t_2$ is a positive constant).

\begin{table}[ht]
\caption{\label{table:averages} \normalsize{Average expansion rates and densities.}}
\begin{center}
\begin{tabular}{@{}lllllll}
\br
  & $t_0(r)$ & $\langle\theta\rangle_{\tau}$ & $\langle\theta\rangle_t$ & $\langle\rho\rangle_{\tau}$ & $\langle\rho\rangle_t$ & $a_{\tau}$ \cr
\mr
  \normalsize{Case 1} & $-r e^{-\frac{2 t_1}{r}}$ & $\frac{2}{\tau}\frac{1}{2}$ & 0 & 0 & 0 & $(\tau/t_2)^{1/3}$ \\[10pt]
  \normalsize{Case 2} & $-r_0 e^{-\frac{2 t_1}{r}} \ln\frac{r+r_0}{r_0}$ & $\frac{2}{\tau}\frac{3\tau+r_0}{3\tau+2 r_0}$ & 0 & $\frac{4 M_{Pl}^2}{3\tau^2}\frac{3\tau}{3\tau+2 r_0}$ & 0 & $[\tau(\tau+2 r_0/3)/t_2^2]^{1/3}$ \\[10pt]
  \normalsize{Case 3} & $\frac{r_0}{1 +r^3/(2 t_1^2 r_0)}$ & $\frac{2}{\tau}$ & $\frac{2}{t}$ & $\frac{4 M_{Pl}^2}{3\tau^2}$ & $\frac{4 M_{Pl}^2}{3 t^2}$ & $(\tau/t_2)^{2/3}$ \\
\br
\end{tabular}
\end{center}
\end{table}

In cases 1 and 2 the big bang occurred at $t=0$ in the center
at $r=0$, while the far away shells of constant $r$
asymptotically approach being infinitely old. In contrast,
in case 3 there is only a finite difference (of
$r_0$) in the big bang time between the center and the
region at asymptotic infinity. The three cases cover the possible
asymptotic behaviours for the quantity $r\tau'$: divergent,
finite and zero.

Scalar quantities (such as the expansion rate) are invariant
under coordinate transformations, but spatial averaging
is not a covariant procedure, and the choice of hypersurface
makes a physical difference. Cases 1 and 2 illustrate the drastic
difference between the hypersurfaces of constant $\tau$ and
constant $t$. In case 1 the average expansion rate on the
hypersurface of constant $\tau$ is 1/2 of the FRW rate,
and in case 2 it starts at 1/2 of the FRW rate and
asymptotically approaches it as $\tau$ grows, and
the scale factor correspondingly interpolates 
between $(\tau/t_1)^{1/3}$ (case 1) and
$(\tau/t_1)^{2/3}$ (the FRW case). However, in both cases 1 and 2
the average expansion rate on the hypersurface of constant
$t$ is zero. 
In case 3 the average over both hypersurfaces yields the
unperturbed expansion rate in terms of the appropriate time.

It may seem strange that the average of the expansion rate is zero,
given that the expansion rate is positive definite. The resolution is
that the average is dominated by the region at asymptotic
infinity, since it contains infinitely more observers than
the region within any given finite radius. Since the age of the region at
large $r$ is asymptotically infinite, its expansion rate is
asymptotically zero, so the average expansion rate goes to zero.
It is for the same reason that the averages turn out to be the
unperturbed values in case 3: the big bang time $t_0$ goes
asymptotically to zero, so there is asymptotically no difference
between $\tau$ and $t$, and therefore no backreaction.
In fact, the averages over the two hypersurfaces are simply
the asymptotic limits of the quantities \re{theta}, \re{rho}
with the appropriate time coordinate held fixed. Since the three
cases we have studied cover all possible limits of the quantity
$r\tau'$, we have exhausted the possibilities for the averages:
any permissible function $t_0$ leads to one of the average
quantities in cases 1 to 3 (apart from a trivial global change
in the normalisation of $t$). The averages over all space do not
depend on the integration measure, since in the asymptotic limit
the measure in the nominator and the denominator cancels.
This feature is peculiar to the spherical symmetry of the LTB model.
In a realistic model of the universe where all points are
statistically equivalent, the integration measure is expected to
make a difference. It would also be more physical to average over
the finite region that has been in causal contact with a given
observer, in which case the backreaction would not depend only
on the asymptotic limit, and would be finite in all cases.

The above examples show that even in the LTB model, where (unlike in the
perturbed FRW universe) the coordinate time $t$ is a physically
measurable quantity, taking averages over the hypersurface of
constant $t$ can give misleading results. We
are interested in the average expansion rate measured
by local observers as a function of the proper time they measure,
so we should compare observers whose clocks show the
same time. Assuming that cosmological observers normalise
their clocks to the locally inferred big bang time $t_0$
instead of synchronising them to some global constant,
observers further out in $r$ will reach the same value of
$\tau$ earlier in terms of $t$: the averaging surface
tilts to the past with increasing $r$.

\section{Discussion}

\paragraph{The Hubble law.}

To get another viewpoint on the backreaction, let us
look at the relation between the average expansion rate
and average density in case 2. From the expressions given in
Table \ref{table:averages} we can solve for $\tau$ in terms of
$\langle\rho\rangle_{\tau}$, and insert this into
the expression for $\langle\theta\rangle_{\tau}$. The
resulting Hubble law is
\bea
  \label{theta1} \langle\theta/3\rangle_{\tau}^2 &=& \frac{4}{9\tau^2} \left(\frac{3\tau+r_0}{3\tau+2 r_0}\right)^2 \\
  \label{theta2} &=& \frac{\langle\rho\rangle_{\tau}}{3 M_{Pl}^2} \left( 1 + \frac{r_0^2}{12 M_{Pl}^2} \langle\rho\rangle_{\tau} \right) \\
  \label{theta3} &=& \frac{4}{9 t_1^2} \frac{1}{a_{\tau}^3} \left( 1 + \frac{r_0^2}{9 t_1^2} \frac{1}{a_{\tau}^3} \right) \ ,
\eea

\noindent where we have written the square of the expansion rate in three
different forms: as a function of the proper time, the average
energy density and the effective scale factor. The FRW relations are
recovered as $r_0/\tau$ goes to zero.

From the first expression, \re{theta1}, it seems that the
expansion is slower than in the homogeneous case.
However, from the second and third expressions, \re{theta2} and
\re{theta3}, it seems as if the universe would instead expand
faster as there is an additional positive
$\langle\rho\rangle_{\tau}^2$-term driving the expansion.
We have the paradoxical-seeming situation that the
expansion is slower than in the FRW case if measured in
terms of the proper time, but faster if measured
in terms of the energy density (or the scale
factor, since conservation of mass implies
$\langle\rho\rangle_{\tau}\propto a_{\tau}^{-3}$).
The reason is that the backreaction reduces (in terms \mbox{of $\tau$)}
$\langle\rho\rangle_{\tau}$ more than it reduces $\langle\theta\rangle_{\tau}$,
as we see from Table \ref{table:averages}. Therefore, the
dependence of $\langle\theta\rangle_{\tau}$ on
$\langle\rho\rangle_{\tau}$ has to be stronger than in
the FRW case to get the dependence on $\tau$ right. This
underlines the point, mentioned in
\cite{Geshnizjani:2002, Rasanen:2003}, that
one should be careful to identify which quantity is
being measured in terms of what parameters when
considering backreaction.

\paragraph{A measure of backreaction.}

In \cite{Rasanen:2003} it was suggested that since the Weyl tensor
gives a measure of the inhomogeneity and anisotropy of a spacetime
it could provide some indication of whether backreaction is important.
For the LTB solution with $E=0$, the ratio of the square of the Weyl
tensor to the square of the scalar curvature is, from \re{metric1} and \re{R},
\bea
  \frac{C_{\alpha\beta\gamma\delta} C^{\alpha\beta\gamma\delta}}{\mathcal{R}^2} = \frac{16}{27} \left(\frac{r\tau'}{\tau}\right)^2 \ ,
\eea

\noindent which does measure the importance of backreaction.
For example, in case 2 it is, when averaged over the
hypersurface of constant $\tau$, essentially
the ratio $(r_0/\tau)^2$. Note that the information entropy
measure introduced in \cite{Hosoya:2004} to quantify the
inhomogeneity of a spacetime measures the difference between
a spacetime and its average, not the difference between the
average and the corresponding smooth spacetime. (At any rate,
the measure was defined for compact domains, and so is
not applicable to the present case.)

One can calculate the backreaction for geometric quantities
other than the expansion rate and the Weyl tensor. One interesting
candidate is the shear, which is zero in the FRW case. Like
the square of the Weyl tensor, the shear turns out to be a
measure of the backreaction both in the LTB case and in the
perturbative case of \cite{Rasanen:2003}.
In the perturbative case, the shear becomes of the order
of the scalar curvature, just like the Weyl tensor. This
is in conflict with observations, but the perturbative
analysis of \cite{Rasanen:2003} is at any rate expected
to break down when backreaction is indicated to be important.

\paragraph{Conclusion.}

We have calculated the average expansion rate and energy density
in the simplest, parabolic, LTB solution analytically for some
example cases (which turn out to cover all possibilities).
Comparison of the averages with each other demonstrates the importance
of the choice of hypersurface, and comparison with the FRW
case shows that backreaction slows down the
expansion if measured in terms of the proper time, but speeds
it up if measured in terms of the energy density or the scale factor.
The calculation is an example of an exact, quantitative
study in backreaction.

Unlike the perturbed FRW metric, the LTB metric is not meant to
be a realistic model of the universe, but it is a useful toy model
for studying backreaction because one can obtain analytical results.
One way  towards a more realistic backreaction calculation
from the first order perturbed FRW model used in
\cite{Rasanen:2003} is to do a consistent second order
perturbative analysis. Another possibility could be
a quasi-perturbative approach where one keeps
to first order perturbation theory for scales which
are in the linear regime, but applies a non-perturbative
solution for scales which have gone non-linear. Since the LTB
model can describe the formation of clusters and voids smoothly
starting from small perturbations \cite{Krasinski}, it is
ideal for this purpose. (A simple version of embedding
LTB solutions into a FRW universe was done in \cite{Kozaki:2002},
and the idea goes back to \cite{Lemaitre:1933}.) This would
involve the LTB solutions with $E\neq0$, and it would be interesting
to first study how the average expansion rate behaves in those
solutions themselves.

\ack

I thank Marco Bruni, Jai-chan Hwang and Dominik Schwarz for discussions,
a number of people including Carsten van de Bruck and Joe Silk for
suggesting looking at the LTB model, and the Helsinki Institute of Physics
for hospitality.

The research has been supported by PPARC grant PPA/G/O/2002/00479,
by a grant from the Magnus Ehrnrooth Foundation and by the
European Union network HPRN-CT-2000-00152,
``Supersymmetry and the Early Universe''.\\

\appendix

\setcounter{section}{1}


\begin{thebibliography}{99}

\bibitem{Pietronero:2004} Pietronero L and Labini F S,
\newblock {\it Statistical physics for complex cosmic structures},
\newblock \brt{astro-ph/0406202}

\bibitem{Ellis} Ellis G F R,
\newblock {\it Relativistic cosmology: its nature, aims and problems}, 1983
\newblock The invited papers of the 10th international conference on general relativity and gravitation, p 215
\nonum Ellis G F R and Stoeger W,
\newblock {\it The 'fitting problem' in cosmology}, 1987
\newblock \CQG{4} 1697

\bibitem{Shirokov:1963} Shirokov M F and Fisher I Z,
\newblock {\it Isotropic Space with Discrete Gravitational-Field Sources. On the Theory of a Nonhomogeneous Universe}, 1963
\newblock {\it Sov. Astron. J.} {\bf 6} 699
\newblock Reprinted in \GRG{30} 1411, 1998

\bibitem{Buchert:1999a} Buchert T,
\newblock {\it On average properties of inhomogeneous fluids in general relativity I: dust cosmologies}, 2000
\newblock \GRG{32} 105
\newblock \brt{gr-qc/9906015}

\bibitem{Buchert:2001} Buchert T,
\newblock {\it On average properties of inhomogeneous fluids in general relativity II: perfect fluid cosmologies}, 2001
\newblock \GRG{33} 1381
\newblock \brt{gr-qc/0102049}

\bibitem{Buchert:1999b} Buchert T, Kerscher M and Sicka C,
\newblock {\it Backreaction of inhomogeneities on the expansion: the evolution of cosmological parameters}, 2000
\newblock \PRD{62} 043525
\newblock \brt{astro-ph/9912347}

\bibitem{Buchert:1995} Buchert T and Ehlers J,
\newblock {\it Averaging inhomogeneous Newtonian cosmologies}, 1997
\newblock Astron. \& Astrophys. {\bf 320} 1
\newblock \brt{astro-ph/9510056}

\bibitem{Mukhanov} Mukhanov V F, Abramo L R W and Brandenberger R H,
\newblock {\it Back Reaction Problem for Gravitational Perturbations}, 1997
\newblock \PRL{78} 1624
\newblock \brt{gr-qc/9609026}
\nonum Abramo L R W, Brandenberger R H and Mukhanov V F,
\newblock {\it The Energy-Momentum Tensor for Cosmological Perturbations}, 1997
\newblock \PRD{56} 3248
\newblock \brt{gr-qc/9704037}

\bibitem{Wetterich:2001} Wetterich C,
\newblock {\it Can Structure Formation Influence the Cosmological Evolution?}, 2003
\newblock \PRD{67} 043513
\newblock\brt{astro-ph/0111166}

\bibitem{Geshnizjani:2002} Geshnizjani G and Brandenberger R,
\newblock {\it Back Reaction And Local Cosmological Expansion Rate}, 2002
\newblock \PRD{66} 123507
\newblock \brt{gr-qc/0204074}

\bibitem{Finelli:2003} Finelli F, Marozzi G, Vacca G P and Venturi G,
\newblock {\it Energy-Momentum Tensor of Cosmological Fluctuations during Inflation}, 2004
\newblock \PRD{69} 123508
\newblock \brt{gr-qc/0310086}

\bibitem{Geshnizjani:2003} Geshnizjani G and Brandenberger R,
\newblock {\it Back Reaction Of Perturbations In Two Scalar Field Inflationary Models}
\newblock \brt{hep-th/0310265}

\bibitem{Rasanen:2003} Rasanen S,
\newblock {\it Dark energy from backreaction},
\newblock 2003 JCAP04(2003)004
\newblock \brt{astro-ph/0311257}

\bibitem{Brandenberger:2004a} Brandenberger R and Mazumdar A,
\newblock {\it Dynamical Relaxation of the Cosmological Constant and Matter Creation in the Universe},
\newblock \brt{hep-th/0402205}

\bibitem{Geshnizjani:2004} Geshnizjani G and Afshordi N,
\newblock {\it Coarse-Grained Back Reaction in Single Scalar Field Driven Inflation}
\newblock \brt{gr-qc/0405117}

\bibitem{Rasanen:2004} Rasanen S,
\newblock {\it Backreaction of linear perturbations and dark energy},
\newblock \brt{astro-ph/0407317}

\bibitem{Brandenberger:2004b} Brandenberger R H and Lam C S,
\newblock {\it Back-Reaction of Cosmological Perturbations in the Infinite Wavelength Approximation}
\newblock \brt{hep-th/0407048}

\bibitem{Tsamis} Tsamis N C and Woodard R P,
\newblock {\it Quantum Gravity Slows Inflation}, 1996
\newblock \NPB{474} 235
\newblock \brt{hep-ph/9602315}
\nonum Tsamis N C and Woodard R P,
\newblock {\it The Quantum Gravitational Back-Reaction on Inflation}, 1997
\newblock {\it Annals Phys.} {\bf 253} 1
\newblock \brt{hep-ph/9602316}
\nonum Abramo L R, Tsamis N C and Woodard R P,
\newblock {\it Cosmological Density Perturbations From A Quantum Gravitational Model Of Inflation}, 1999
\newblock {\it Fortsch. Phys.} {\bf 47} 389
\newblock \brt{astro-ph/9803172}
\nonum Woodard, R P,
\newblock {\it Effective Field Equations of the Quantum Gravitational Back-Reaction on Inflation}
\newblock \brt{astro-ph/0111462}

\bibitem{Brandenberger:2000} Brandenberger R H,
\newblock {\it Back Reaction of Cosmological Perturbations},
\newblock{hep-th/0004016}

\bibitem{Brandenberger:2002} Brandenberger R H,
\newblock {\it Back Reaction of Cosmological Perturbations and the Cosmological Constant Problem},
\newblock \brt{hep-th/0210165}

\bibitem{Schwarz:2002} Schwarz D J,
\newblock {\it Accelerated expansion without dark energy}
\newblock \brt{astro-ph/0209584}

\bibitem{Unruh:1998} Unruh W,
\newblock {\it Cosmological long wavelength perturbations},
\newblock \brt{astro-ph/9802323}

\bibitem{Lemaitre:1933} Lema\^{\i}tre A G,
\newblock {\it The Expanding Universe}, 1933
\newblock {\it Ann. Soc. Sci Bruxelles} {\bf A53} 51 (in French)
\newblock Reprinted in 1997 \GRG{29} 641

\bibitem{Tolman:1934} Tolman R C,
\newblock {\it Effect of Inhomogeneity on Cosmological Models}, 1934
\newblock {\it Proc. Nat. Acad. Sci. USA} {\bf 20} 169
\newblock Reprinted in 1997 \GRG{29} 935

\bibitem{Bondi:1947} Bondi H,
\newblock {\it Spherically symmetrical models in general relativity}, 1947
\newblock \MNRAS{107} 410

\bibitem{Krasinski:1997} Krasi\'{n}ski A,
\newblock {\it Inhomogeneous Cosmological Models}, 1997
\newblock Cambridge University Press, Cambridge

\bibitem{Sussman:2001} Sussman R A and Trujillo L G,
\newblock {\it New variables for the Lema\^{\i}tre-Tolman-Bondi dust solutions}, 2002
\newblock \CQG{19} 2897
\newblock \brt{gr-qc/0105081}

\bibitem{Krasinski} Krasi\'{n}ski A and Hellaby C,
\newblock {\it Structure formation in the Lema\^{\i}tre-Tolman model}, 2002
\newblock \PRD{65} 023501
\newblock \brt{gr-qc/0106096}
\nonum \dash
\newblock {\it More examples of structure formation in the Lema\^{\i}tre-Tolman model}, 2004
\newblock \PRD{69} 023502
\newblock \brt{gr-qc/0303016}

\bibitem{clusters} Fouqu\'e P, Solanes J M, Sanchis T and Balkowski C,
\newblock {\it Structure, mass and distance of the Virgo cluster from a Tolman-Bondi model}, 2001
\newblock \AA{375} 770
\newblock \brt{astro-ph/0106261}
\nonum Hanski M, Theureau G, Ekholm T and Teerikorpi P,
\newblock {\it Kinematics of the local universe IX. The Perseus-Pisces supercluster and the Tolman-Bondi model}, 2001
\newblock \AA{378} 345
\newblock \brt{astro-ph/0109080}

\bibitem{Hellaby:1988} Hellaby, C,
\newblock {\it Volume Matching in Tolman Models}, 1988
\newblock \GRG{20} 1203

\bibitem{Mustapha:1997} Mustapha N, Bassett B A, Hellaby C and Ellis G F R,
\newblock {\it Shrinking II -- The Distortion of the Area Distance-Redshift Relation in Inhomogeneous Isotropic Universes}, 1998
\newblock \CQG{15} 2363
\newblock \brt{gr-qc/9708043}

\bibitem{Buchert:2003} Buchert T and Carfora M,
\newblock {\it The Cosmic Quartet - Cosmological Parameters of a Smoothed Inhomogeneous Spacetime},
\newblock \brt{astro-ph/0312621}

\bibitem{Hellaby:1985} Hellaby C and Lake K,
\newblock {\it Shell crossings and the Tolman model}, 1985
\newblock \APJ{290} 381
\newblock Erratum in 1986 \APJ{300} 461

\bibitem{Misner:1970} Misner C W, Thorne K S and Wheeler J A,
\newblock {\it Gravitation}, 1973
\newblock W.H. Freeman and Company, New York,
\newblock p 505

\bibitem{Hosoya:2004} Hosoya A, Buchert T and Morita M,
\newblock {\it Information Entropy in Cosmology}, 2004
\newblock \PRL{92} 141302
\newblock \brt{gr-qc/0402076}

\bibitem{Kozaki:2002} Kozaki H and Nakao K,
\newblock {\it Volume Expansion of Swiss-Cheese Universe}, 2002
\newblock \PRD{66} 104008
\newblock \brt{gr-qc/0208091}

\end{thebibliography}
\end{document}